\documentstyle[pre,aps,multicol]{revtex}
\begin{document}
\draft
\title{Persistence, Poisoning, and Autocorrelations in Dilute Coarsening}
\author{Benjamin P. Lee}
\address{Polymers Division, National Institute of Standards and Technology,
Gaithersburg, MD 20899, USA}
\author{Andrew D. Rutenberg}
\address{Theoretical Physics, University of Oxford, Oxford OX1 3NP, United
Kingdom }
\date{Phys.\ Rev.\ Lett.\ {\bf 79}, 4842 (1997) -- Received 29 July 1997}
\maketitle
\begin{abstract}
We calculate the exact autocorrelation exponent $\lambda$ and 
persistence exponent $\theta$, and also amplitudes, 
in the dilute limit of phase ordering for dimensions $d\geq 2$.
In the Lifshitz-Slyozov-Wagner limit of
conserved order parameter dynamics we find $\theta=\gamma_d\epsilon$,
a universal constant times the volume fraction. For autocorrelations, 
$\lambda=d$ at intermediate times, with 
a late time crossover to $\lambda \geq d/2+2$.
We also derive $\lambda$ and $\theta$ for globally conserved
dynamics and relate these to
the $q \to \infty$-state Potts model and soap froths, proposing
new poisoning exponents.
\end{abstract}
\kern -0.05truein
%%%%%%%%%%%%%%%%%%%%%%%%%%%%%%%%%%%%%%%%%%%%%%%%%%%%%%%%%%%%%%%%%%%%%%%%%%%
%
\kern -0.15truein
\begin{multicols}{2}
While much has been learned about the coarsening kinetics
that follows a temperature quench from a single- to a multi-phase state 
\cite{Bray94}, relatively little has been
established for certain recently introduced dynamical exponents.
It is accepted that the characteristic length scale of strongly correlated 
regions grows as a power law in time, $L \sim t^{1/z}$, with universal $z$ 
\cite{Bray94}.  Most non-conserved order parameter systems yield $z=2$
and those with scalar conserved order parameter dynamics yield $z=3$,
{\em independent\/} of the system dimensionality 
$d$ or of conserved quantities such as the volume fraction $\epsilon$ of the 
minority phase.  Consequently, persistence 
\cite{Marcos-Martin95,Bray94b,COM:global,Derrida95,Cardy95,%
Majumdar96,Majumdar96c,Levitan96,Derrida97,Yurke96}
and autocorrelation 
\cite{Fisher88,Mason93,Bray95,Sire95,Yeung96,Marko95}
exponents, $\theta$ and $\lambda$ respectively,
are being explored in the hope that they contribute to a 
characteristic set of universal exponents analogous to those of equilibrium 
criticality.  However, fundamental questions remain
about the universality of these new exponents, and even of 
the existence of power laws in the relevant quantities.

The persistence exponent $\theta$,
introduced in the experimental study of breath figures 
\cite{Marcos-Martin95}, 
is defined by the power-law decay, $P(t_1,t_2)\sim t_2^{-\theta}$, of the
probability that a stochastic variable has not crossed some threshold ---
typically its mean --- between the times $t_1$ and $t_2$
\cite{Bray94b}.  We consider
the persistence of a {\em local\/}, scalar order parameter $\phi({\bf r},t)$ 
(rescaled so that in equilibrium $\phi=\pm 1$), given
by the fraction of the system that has not undergone phase change 
between $t_1$ and $t_2$ \cite{COM:global}.  

In $d=1$, 
$\theta$ has been calculated exactly for the non-conserved
$q$-state Potts model \cite{Derrida95}, and has been
shown to be universal by renormalization group methods in the Ising ($q=2$) 
case \cite{Cardy95}.
In higher dimensions, studies 
have focussed on diffusion models \cite{Majumdar96}, which exhibit 
non-trivial values of $\theta$, and
on related Gaussian approximations for non-conserved ordering kinetics 
\cite{Majumdar96,Majumdar96c}.
For $d=2$, these approximate results
compare well with simulations \cite{Levitan96,Derrida97} 
and with experiments on twisted nematic liquid crystals \cite{Yurke96}, 
an Ising analog.  However, there have been no
previous studies of persistence for conserved coarsening dynamics.

In phase-ordering systems the autocorrelation function,
$A(t_1,t_2) \equiv \langle \phi({\bf r},t_1) \phi({\bf r},t_2) 
\rangle-\langle\phi\rangle^2$,
decays asymptotically as $A(t_1,t_2) \sim [L(t_1)/L(t_2)]^\lambda$, which
defines $\lambda$ \cite{Bray94,Fisher88}.  For non-conserved
scalar coarsening, $\lambda$ has been measured experimentally in $d=2$
\cite{Mason93} and calculated in $d=1$ \cite{Bray95}.  
Approximate calculations and numerical results have been obtained
for various $\epsilon$ and $d$ \cite{Sire95}
in the case of {\em globally\/} conserved dynamics
--- i.e., a non-conserved order parameter (hence $z=2$) subject to
a constraining field that maintains the total volume fraction of each phase.
{\em Locally\/} conserved dynamics has been studied numerically in two
\cite{Yeung96,Marko95} and three \cite{Marko95} dimensions, and a
formal asymptotic bound, 
$\lambda\geq d/2+2$,  has been established by Yeung, Rao and 
Desai (YRD) \cite{Yeung96}, but otherwise little theoretical progress has been 
made.

In this Letter, we study persistence and autocorrelations for both 
locally and globally conserved dynamics.
We focus on the asymptotic, late stage regime which follows a quench to a
subcritical temperature $T<T_c$, in the limit of vanishing volume fraction,  
$\epsilon\to 0^+$.  
This is the limit of the classic Lifshitz-Slyozov-Wagner (LSW) theory,
which firmly established $z=3$ for dilute, locally conserved coarsening
\cite{Bray94,LSW61}. We use LSW theory in a similar spirit, to obtain 
$\theta$ and $\lambda$.  A summary of our results follows.

For locally conserved dynamics we compute $\theta$ for all $d\geq 2$, and
demonstrate that 
(i) the persistence
decays as a power law, $P(t_1,t_2) \sim  (t_1/t_2)^\theta$,
(ii) the exponent is a function of the volume fraction, going as
\begin{equation}\label{EQN:thetaLSW}
	\theta = \gamma_d\epsilon
\end{equation}
in the small $\epsilon$ limit, and 
(iii) $\gamma_d$ is universal in that it does not depend on the surface 
tension, quench depth, temperature, or mobility, with  
$\gamma_2 \simeq 0.39008$, and $\gamma_3 \simeq 0.50945$.   
A large-$d$ expansion gives the asymptotic series
\begin{equation}\label{EQN:larged}
\gamma_d  = \sqrt{3 d/8\pi}[1 + \sum_{m=1}^k a_m d^{-m} +O(d^{-k-1})],
\end{equation}
which is quite accurate in $d=2,3$ when truncated at $k=3$ \cite{series,long}.
We also compute the order $\epsilon^{3/2}$ 
corrections to (\ref{EQN:thetaLSW}) in $d=3$.

For the autocorrelation function we find $\lambda=d$
as $\epsilon\to 0^+$, with explicit, universal expressions for both the 
amplitude {\em and\/} logarithmic corrections. We also
present a physical scaling argument that predicts a crossover to 
$\lambda \geq d/2+2$ for any finite $\epsilon$, after
$t_2 \gtrsim \epsilon^{-3/d} t_1$, thus satisfying the YRD bound 
\cite{Yeung96}.

Next, we consider globally conserved (GC) dynamics, again in the
small $\epsilon$ limit, and find
$\theta$ to have the same form (\ref{EQN:thetaLSW}), with a
different, universal $\gamma_d^{GC}$.  In particular,
$\gamma_2^{GC} \simeq 0.48797$ and $\gamma_3^{GC} \simeq 0.62450$.
In large $d$  
\begin{equation}\label{EQN:gammaGC}
	\gamma_d^{GC}=\sqrt{d/2\pi} \> [1-{\textstyle{1\over 3}}d^{-1}
	+{\textstyle {43 \over 288}}d^{-2} + O(d^{-3})],
\end{equation}
which is highly accurate for all $d$ \cite{long}.  The large $d$
asymptote, $\theta \simeq 0.40\epsilon\sqrt d$ is similar to the
approximate result $\theta \simeq 0.15\sqrt d$ obtained at $\epsilon={1\over 2}$
\cite{Majumdar96} (where the GC dynamics is equivalent to
non-conserved).  The autocorrelation 
exponent is $\lambda=d$ \cite{Sire95}, with no crossover expected
at late times. We also find universal amplitudes {\em and\/} logarithmic
corrections. 

Finally, we draw a connection between the above results for 
{\em persistence\/} in GC dynamics with {\it poisoning\/} (defined
below) in soap froths. First, there is 
some evidence, mainly in $d=2$, that soap froths have the 
same asymptotic dynamics as the non-conserved $q$-state Potts model in 
the $q \to \infty$ limit \cite{Grest88}.  Second, the coarsening of the
$q$-state Potts model and that of GC dynamics with 
$\epsilon=1/q$ were shown to be equivalent as $q \to \infty$, within 
a Gaussian approximation scheme \cite{Sire95}.
However, the details of the topological rearrangements may be different 
between Potts models and soap froths \cite{Grest88}, and further, while
both Potts models and soap-froths have vertices, GC systems do not. 

Nevertheless, numerical studies found little difference in $\lambda$ between 
the Potts and GC models \cite{Sire95}, implying that they may
lie in the same dynamic universality class.  Also, the Potts persistence 
exponent, which measures the volume never visited by a wall, is given 
via the GC correspondence by $\theta=d/2$ as $q\to\infty$ \cite{BenNaim97},
consistent with Potts simulations \cite{Levitan96,Hennecke97} and with 
experiments on $d=2$ soap froths \cite{Tam97}.  To further explore these 
analogies, we define a new set of {\em poisoning\/} exponents, 
$\theta_{\Sigma}$, that give the decay of volume that has 
never been visited by any of a {\em set\/} of phases that 
occupy a {\em total\/} volume fraction $\Sigma$.  This poisoning should 
provide a more delicate test of the underlying dynamics 
than autocorrelations or persistence, and
may be directly explored via simulations of Potts models and experiments
on foams \cite{label}.  By use of the GC correspondence we obtain,
for $\Sigma \ll 1$,
\begin{equation}\label{EQN:Sigma}
	\theta_{\Sigma}=\gamma_d^{GC} \Sigma.
\end{equation}

We begin with LSW theory, which applies to widely separated 
drops of minority phase as
$\epsilon\to 0^+$.  This theory provides the only solution of a 
phase-ordering system in $d>1$ with topological defects ---  
in this case with domain walls.  Drops of radius $R$ evolve according to
 $R^2\dot R=\alpha_d (R/R_c-1)$, where $\alpha_d$ is a nonuniversal 
constant \cite{Bray94}, and the dot indicates a time-derivative.  
Here $R_c= ({4\over 9}\alpha_d t)^{1/3} \sim L$ is the critical
radius, where drops shrink for $R< R_c$ and grow for $R>R_c$.
The density $n(R,t)$ of droplets of size $R$ at time 
$t$ obeys a continuity equation $\dot{n} =-\partial_R(\dot R n)$, which
leads to a scaling solution $n_d(R,t)=R_c(t)^{-d-1}f_d(R/R_c)$, with 
\begin{equation}\label{EQN:distribution}
	f_d(x) = 
	{\epsilon F_d x^2 \exp[-d/( 3-2x)]\over (3+x)^{1+4d/9}
	(3-2x)^{2+5d/9}},
\end{equation}
and $f_d=0$ for $x \geq 3/2$ \cite{Bray94,d2,Yao93}.
The normalization constant $F_d$ is determined by the volume fraction of 
the minority phase, $\epsilon = V_d\int_0^\infty dx f_d(x) \, x^d$,
where $V_d \equiv\pi^{d/2}/\Gamma(1+{1\over 2}d)$ is the unit 
$d$-sphere 
volume.  This gives $F_2 \simeq 37.752$ and $F_3 \simeq 186.13$, and the
large-$d$ expansion $V_d F_d = e^d 2^{1/2+8d/9} 
\sqrt{27 d/\pi} [1+\sum_{m=1}^k a_m d^{-m} +O(d^{-k-1})]$ \cite{series}.
The total number density of drops can be shown to be
$n(t) =\epsilon F_d/(4d3^de^{d/3})[R_c(t)]^{-d}$.

The droplet growth equation can be written in terms of the scaled 
size $x\equiv R/R_c(t)$ as
$3 t x^2\dot x =- (x+3)(x-{\textstyle{3\over 2}})^2$; where
$\dot x<0$ for all $x \geq 0$. This
may be integrated to give the trajectory
\begin{equation}\label{EQN:trajectory}
	t_f=t (1+{\textstyle{1\over 3}}x)^{4/3}
	(1-{\textstyle{2\over 3}}x)^{5/3}
	\exp\left[2x/( 3-2x) \right],
\end{equation}
which is parameterized by the time of complete evaporation $t_f$.  From 
(\ref{EQN:trajectory}), we find an expansion for $t_1 \ll t_2 \leq t_f$
which will prove useful below:
\begin{equation}
\label{EQN:dropsize}
	x(t_1) = {\textstyle{3\over 2}}\left[ 1-\delta(t_1/t_2) 
	+ \ldots \right],
\end{equation}
where 
$\delta(t_1/t_2) = 1/\left[ \ln(t_2/t_1) +{5\over 3}\ln\ln(t_2/t_1) \right]$. 
This leading correction to $x(t_1)$, 
related to the essential singularity in $f_d(x)$ 
at $x={3\over 2}$, is {\em universal\/} and independent of  $x(t_2)$.

Using these LSW results, we can 
calculate $P^<(t_1,t_2)$, the persistent or unpoisoned volume fraction 
of minority phase, and $P^>(t_1,t_2)$, that of the majority phase.  
The total persistent volume $P=P^>+P^<$ will decay with the slower of the 
two unpoisoned fractions.

Only  droplets that have survived to time $t_2$ contribute to $P^<$,
the unpoisoned {\it minority\/} phase, and their density decays as
$n\sim R_2^{-d} \sim t_2^{-d/3}$, using the notation $R_i \equiv R_c(t_i)$.  
The density of droplets that have survived
and are {\em smaller\/} at $t_2$ than they were at $t_1$ is
\begin{equation}
	R_2^{-d}\int_0^{x_1(t_1/t_2)^{1/3}}dx f_d(x) \sim 
	R_2^{-d-3} \sim t_2^{-d/3-1},
\label{EQN:smaller}
\end{equation}
where we have used the small $x$ behavior $f_d(x)\sim x^2$ 
for $t_2 \gg t_1$, and $R_c \sim t^{1/3}$.  Hence, droplets that have shrunk
comprise a vanishing fraction of those surviving at $t_2$, so 
that no surviving droplets have been poisoned at time $t_2 \gg t_1$.
Consequently, $P^<$ is asymptotically the number 
density $n(t_2)$ times the volume at the initial time
$t_1$ of those largest droplets, 
$V_d[x(t_1)R_1]^d$.  Using $x(t_1)$ from (\ref{EQN:dropsize}),
\begin{equation}
\label{EQN:Pminority}
	P^<(t_1,t_2) = {\textstyle{1\over 4}}\epsilon B_d(t_1/t_2)^{d/3}
	[1-d\delta(t_1/t_2)+\ldots],
\end{equation}
where  $ B_d=F_d V_d/ ( 2^d d e^{d/3} )$, with $B_2 \simeq 7.6115$ and 
$B_3 \simeq 11.951$, {\em and\/} the leading logarithmic corrections 
$\delta(t_1/t_2)$ are universal.

The {\em majority\/} phase, with unpoisoned volume fraction $P^>$, 
can only be poisoned by growing drops, i.e.\ those with  $R>R_c$.
Since the drop positions are uncorrelated in the dilute 
limit \cite{LSW61} --- the key feature which makes LSW theory soluble
--- it follows that the unpoisoned regions must be
uncorrelated as well, leading to
\begin{equation}\label{EQN:Pevolution}
	\partial_t {P}^>(t_1,t)=
	- \dot{v}(t) P^>(t_1,t),
\end{equation}
where $\dot{v}(t)$ is the rate of encroachment by
minority phase.  From LSW theory we have
\begin{equation}
\label{EQN:encroachment}
	\dot{v}(t) = \int_{R_c}^{R_{\rm max}} dR \, V_d R^d \partial_t n(R,t)
       		= V_d R_c^d n(R_c,t) \dot R_c,
\end{equation}
where the second equality comes from mass conservation of drops larger than
$R_c(t)$, i.e.\ $\partial_t \int_1^{3/2} dx \, x^d f_d(x)=0$.  Using 
$R_c \propto t^{1/3}$, we obtain $\dot{v}= V_d f_d(1)/(3t) \equiv \epsilon 
\gamma_d/t$ where 
\begin{equation}
	\gamma_d = F_d V_d /( 4^{1+4d/9} 3 e^d )
\end{equation} 
is a universal constant.  Combining (\ref{EQN:Pevolution}) and
the initial condition, $P^>(t_1,t_1)=1-\epsilon$, we find
\begin{equation}\label{EQN:Pmajority}
	P^>(t_1,t_2)=(1-\epsilon)\left(t_1 / t_2\right)^{\gamma_d\epsilon},
\end{equation}
so $P^>$ indeed decays as a power law.  Remarkably, this result is 
valid for all $t_2>t_1$ in the scaling regime, not just when $t_2\gg t_1$.
As expected, $P^>$ decays slower than $P^<$ , and so 
$P(t_1,t_2)\sim P^>(t_1,t_2)$,
leading to equation (\ref{EQN:thetaLSW}) for $\theta$.

In order to derive $\lambda$ from the persistence, it is convenient to change 
field variables to $\psi=(\phi+1)/2$
(with minority phase $\psi=1$ and majority $\psi=0$), 
giving $\langle\psi\rangle=\epsilon$ 
and the autocorrelation function
$A(t_1,t_2)= 4[\langle\psi({\bf r},t_1)\psi({\bf r},t_2)\rangle-\epsilon^2]$.
The two-time average is then the probability of finding a given point 
inside minority droplets at both $t_1$ and $t_2$.  The contribution from
unpoisoned regions is exactly $P^<(t_1,t_2)$, whereas poisoned 
regions that find themselves in a droplet again at $t_2$ contribute
$[\epsilon-P^<(t_1,t_2)]^2$.  To leading order in $\epsilon$, 
\begin{equation}\label{EQN:autocorrelation}
     	A(t_1,t_2) = \epsilon B_d (R_1 / R_2)^d 
	[1-d\delta(t_1/t_2)+\ldots],
\end{equation}
giving $\lambda=d$ in the dilute limit (this was noted before 
in the GC case \cite{Sire95}). This exponent depends solely
on the existence of a scaling distribution of uncorrelated drops;
in the LSW limit the details of the drop distribution and evolution serve only
to determine the universal amplitude and leading 
logarithmic corrections.

We expect (\ref{EQN:autocorrelation}) to hold for $R_2$ much less 
than the drop separation at $t_1$, 
$\widetilde R_1 \sim \epsilon^{-1/d} R_1$.  
For $\epsilon \to 0^+$ this is forever. Below we discuss 
$\epsilon>0$, where the drop separation is finite and correlations
must be considered. 

We turn now to globally conserved dynamics, where the analog of LSW theory
was given by  Sire and Majumdar \cite{Sire95}. Droplet growth 
follows $\dot{R} = \alpha_d[1/R_c(t)-1/R]$,
where $R_c(t)$ represents a time-dependent applied field
tuned to maintain the volume fraction, $\epsilon$.  
Combining the droplet growth with the continuity equation,
as in LSW,  we find a scaling solution for the droplet density,
$n(R,t)=R_c^{-d-1}f^{GC}(R/R_c)$, when $R_c(t)=({1\over 2} \alpha_dt)^{1/2}$,
with the distribution
\begin{equation}
	f^{GC}_d(x)=\epsilon F_d^{GC}x(2-x)^{-d-2}\exp[-2d/(2-x)]
\end{equation}
for $x< 2$, and $f_d=0$ otherwise.  The normalization condition,
$\epsilon = V_d \int_0^\infty dx \, x^d f_d^{GC}(x)$, determines $F_d^{GC}$,
with $F_2^{GC} \simeq 16.961$ and $F_3^{GC} \simeq 120.29$.  
A large-$d$ expansion yields the 
excellent approximation $V_d F_d^{GC} = e^{2d}\sqrt{2d/\pi}
\left[ 1-{1\over 3}d^{-1} +{43\over 288}d^{-2}  - {\textstyle{1033\over
	25920}}d^{-3} + O(d^{-4})\right]$. 
The droplet density is
$n^{GC}(t)=\epsilon F_d^{GC}(2e)^{-d}/d$ \cite{long}.
Integrating the scaled growth equation,
$2 t x \dot{x} = -(2-x)^2$, gives the trajectory
$t_f = t \, (1-{\textstyle{1\over 2}}x)^2  \exp[2x/(2-x)]$.
Hence, drops surviving to time $t_2 \gg t_1$ have
\begin{equation}
x(t_1)=2[1-\delta^{GC}(t_1/t_2)+ \ldots],
\end{equation}
where the leading correction is 
$\delta^{GC}(t_1/t_2) = 2 / \left[\ln(t_2/t_1) + 2 \ln \ln(t_2/t_1) \right]$.
The density of drops that are 
smaller at $t_2$ than at $t_1$ decays as $R_2^{-d-2}$, so these are 
again negligible asymptotically.  Consequently, the autocorrelation function 
$A(t_1,t_2)=4P^<(t_1,t_2)$  is
\begin{equation}
	A(t_1,t_2) = \epsilon B_d^{GC} (R_1 / R_2)^d 
	[1- d\delta^{GC}(t_1/t_2) + \ldots],
\end{equation}
with $B_d^{GC}=V_d F_d^{GC}/(d e^d)$, giving $B_2^{GC} \simeq 3.6057$, 
$B_3^{GC}\simeq 8.3623$. The leading 
logarithmic corrections in $\delta^{GC}(t_1/t_2)$ are universal.

The calculation of majority poisoning and persistence goes through the same
as before, using (\ref{EQN:encroachment}) and (\ref{EQN:Pevolution}), with 
the result $\theta=\gamma_d^{GC}\epsilon$.  The different growth exponent 
$R_c\sim t^{1/2}$ gives  $\gamma_d^{GC}=V_d f_d^{GC}(1)/2\epsilon =
V_d F_d^{GC} e^{-2d}/2$.  The result is (\ref{EQN:gammaGC}).

Up to this point, calculations of, say, $f_d(x)$, have been for the 
leading $O(\epsilon)$ term, for which the drops may be regarded as 
uncorrelated.
Higher order effects such as droplet collisions and diffusion-mediated 
interactions will lead to correlations in the drop sizes and positions.  
However, screening of the diffusion field has been shown in $d=3$
to contribute $O(\epsilon^{3/2})$ corrections for {\em uncorrelated\/} drops
\cite{Yao93,Marqusee84,Tokuyama93}, which is thus believed to
represent the leading correction to LSW theory, with correlations coming 
in only at $O(\epsilon^2)$ \cite{Marqusee84}.  Note that with GC
dynamics there is no diffusion field, hence the leading 
corrections, due to collisions, are expected to be of $O(\epsilon^2)$.

With our existing machinery, then, we can compute the leading
corrections to the LSW exponents.  Since drops are
uncorrelated to $O(\epsilon^{3/2})$, $\lambda=d$ remains unchanged.
However, the distribution function is 
$\tilde f_3(x) = f_3(x)[1 + \epsilon^{1/2} \{ G_3 + g_3(x) \}]+O(\epsilon^2)$,
with
\begin{equation}
	g_3(x) = 
	b_0\biggl[ 2 \ln\biggl({3+x\over {\textstyle{3\over 2}}-x}\biggr)
	+{14\over x+3}+{64 x - 87\over 4(3-2x)^2} \biggr]
\end{equation}
where $b_0={1\over 9}\sqrt{\pi F_3/e}\simeq 1.6297$, and  we have maintained 
$x\equiv R/R_c$ \cite{long}.  
The normalization condition determines $G_3 \simeq -3.4047$.
This leads, via (\ref{EQN:encroachment}), to the persistence exponent 
$\theta \simeq 0.50945 \epsilon - 0.14969 \epsilon^{3/2} +O(\epsilon^2)$.

The leading correction to $\lambda$ for $\epsilon > 0$ is of a different
nature. At sufficiently late times $t_2$, 
$\lambda$ becomes strongly dependent on correlation effects \cite{long}.  
This occurs when drops grow to be larger than their 
earlier spacing, $R_2 \gtrsim \widetilde R_1 \sim \epsilon^{-1/d} R_1$.
Then, the autocorrelation function is no longer given by the unpoisoned 
minority volume fraction,
since each drop at $t_2$ covers {\em many\/} drops from $t_1$. As a result,
the decay of autocorrelations becomes dominated by the fluctuations
in drop density at $t_1$, which are described by the correlations.
To see this, we use the small-$k$ behavior of the structure, 
$S(k,t)\equiv \langle \psi_k \psi_{-k} \rangle \sim  R_c^{d+4} k^4$
\cite{Yeung88,Fratzl91}.  At time $t_1$, the fluctuations $\delta V_\psi$ 
in the volume covered by $\psi=1$ within an region of size $R_2$ is
(up to numerical factors) $\delta V_\psi^2 \sim \int d^dk \int_0^{R_2} d^dr
\int_0^{R_2} d^d r' e^{-i k r'} S(k,t_1) \sim R_2^{2d} \int_0^{1/R_2} d^dk
S(k,t_1)$, thus  $\delta V_\psi \sim \pm R_1^{(d+4)/2} R_2^{(d-4)/2}$.   
In the case that the drops at $t_2$ coincide with positive fluctuations 
in the drop density 
at $t_1$, then this volume $\delta V_\psi$ will be contributed to $A(t_1,t_2)$
for every drop at $t_2$, of number density $1/R_2^d$. 
This gives the autocorrelation decay
$A(t_1,t_2) \sim  (R_1/R_2)^{(d+4)/2}$, which 
saturates the YRD bound \cite{Yeung96}. In the case of weaker
correlations between drops at $t_2$ and fluctuations in drop density
at $t_1$, the autocorrelations decay faster. With both cases, we
recover the YRD bound $\lambda \geq d/2+2$.  Comparison with the 
$\epsilon \to 0^+$ result $\lambda=d$, we see that, at least for $d<4$, 
the asymptotic decay of autocorrelations becomes faster when 
correlations suppress fluctuations in the drop density at large length-scales.

In summary, we have demonstrated the existence of power laws in the
autocorrelations, persistence and poisoning of  LSW and GC systems for
$T<T_c$ coarsening, and 
have calculated the exact asymptotic amplitudes and exponents.  Our results
are universal for isotropic systems such as polymer blends (before 
hydrodynamic regimes).  While the exponent $\theta$ is small in the
dilute regime, it should be measurable
since (\ref{EQN:Pmajority}) holds for all $t_2 > t_1$.

In the future, we hope that the poisoning exponents $\theta_\Sigma$ are
measured for foams and large-$q$ Potts models as a sensitive probe of 
their equivalence, and compared with (\ref{EQN:Sigma}). It would also be 
worthwhile to extend the independent interval approximation
\cite{Majumdar96} to the dilute globally-conserved case, to 
further explore the common $\gamma_d \sim \sqrt{d}$ asymptote.  
Finally, our late-time 
crossover and logarithmic corrections in the autocorrelations
may survive in larger filling fractions. Studying $\lambda$ and its 
crossover as a function of $\epsilon$ may resolve the current ambiguities 
\cite{Yeung96,Marko95} about autocorrelation decay in conserved systems. 

%%%%%%%%%%%%%%%%%%%%%%%%%%%%%%%%%%%%%%%%%%%%%%%%%%%%%%%%%%%%%%%%%%%%%%%%%%%
We thank Alan Bray and Claude Godr\`eche for discussions.  BPL was supported 
by an NRC Research Associateship; ADR by EPSRC grant GR/J78044.
%%%%%%%%%%%%%%%%%%%%%%%%%%%%%%%%%%%%%%%%%%%%%%%%%%%%%%%%%%%%%%%%%%%%%%%%%%%

\kern -0.2truein

\end{multicols}

\begin{references}
\bibitem{label} In a foam the labeling might be done physically, or it
	could be done virtually with real time imaging.
\bibitem{d2} Our results apply in $d=2$ as well, 
	where $\alpha_2 \sim 1/\ln(\epsilon^{-1/2})$ for
	small $\epsilon$ \cite{Yao93}.
\end{references}
\end{document}